\documentclass[pre,aps,draft,showpacs]{revtex4}

\usepackage[dvips,final]{graphicx}
\usepackage{bm}
\newcommand{\utensor}{\underline{\underline{u}}}
\newcommand{\vtensor}{\underline{\underline{v}}}

\newcommand{\Tr}[1]{\text{Tr}\underline{\underline{#1}}}
\newcommand{\vtr}{\tilde{\underline{\underline{v}}}}
\newcommand{\Q}{\underline{\underline{Q}}}
\newcommand{\vtw}{\tilde{v}}
\newcommand{\onethird}{\frac{1}{3}}

\begin{document}

\title{The isotropic--cholesteric transition in liquid--crystalline gels}
\author{ Robert A. Pelcovits}
\affiliation{Department of Physics, Brown University, Providence, Rhode Island 02912}
\author{Robert B. Meyer}
\affiliation{The Martin Fisher School of Physics, Brandeis University, Waltham, Massachusetts 02254-9110}
\date{\today}

\begin{abstract}

In a nematic gel, the appearance of nematic order is accompanied
by a spontaneous elongation of the gel parallel to the nematic
director.  If such a gel is made chiral, it has a tendency to form
a cholesteric helical texture, in which local elongation of the
gel parallel to the nematic director is suppressed due to the
requirement of elastic compatibility.  We show that a conical
helix in which the director makes an oblique angle with respect to
the helix axis serves as an energy minimizing compromise between
the competing tendencies for elongation and twisting.  We find the
dependence of the helical cone angle and pitch on the strength of
the chirality, and determine the change in sample shape at the
isotropic to cholesteric phase transition.

 \end{abstract}
\maketitle

\section{\label{sec:Introduction}Introduction}

A nematic gel is basically a polymer gel embedded in a nematic
solvent \cite{Warner and Terentjev:96}.  To make these two systems
compatible, the polymer of the gel is normally formed from monomer
molecules that have a nematogen-like side chain attached to the
reactive group that will form the polymer backbone.  As in an
ordinary polymer gel, the volume fraction of nematic solvent is of
order 90\%. The fundamental interaction between the polymer chains
and the nematic is an orientational coupling.  Here we will assume
that the flexible  polymer backbone has a tendency to be aligned
parallel to the nematic director.  The primary implication of this
coupling is that at the isotropic to nematic phase transition, a
typical polymer coil in the gel transforms from a spherical object
in the isotropic phase to a prolate ellipsoid aligned with its
long axis parallel to the director in the nematic phase.  Because
the polymer coils are cross-linked in the gel, this shape
transformation carries over from the individual polymer coils to
the sample as a whole, leading to a spontaneous elongation of the
entire sample parallel to the director.

This simple view of the transition assumes that the director is
uniformly aligned throughout the sample.  Experimentally this is
found not to be the case in real materials, unless some step has
been taken to bias the orientation of the director in the material
\cite{Finkelmann:91,Mitchell:93}. In an unbiased sample,
some kind of macroscopically isotropic, multidomain, nematic
appears, the exact nature of which is not currently known. Since
spontaneous elongation parallel to the director minimizes the free
energy, this multi-domain state must combine some degree of local
elongation, along with overall average isotropy. The problem is to
see how to combine local elongations of many different
orientations in an elastically compatible way.

In this paper, we study a problem that has elements of the general
problem alluded to above.  We examine the question of how a single
domain of a cholesteric mesophase can form at the isotropic to
cholesteric transition.  In the ordinary cholesteric helix, the
underlying nematic director is oriented perpendicular to the helix
axis.  If one tries to imagine local spontaneous elongations
parallel to the director, there is a serious problem, since every
quarter turn of the helix, the elongations would be in orthogonal
directions, a clear impossibility.  We suggest as a possible
solution to this problem a compromise between the full twisting of
the cholesteric helix, and the overall elongation of the sample.
It is a conical helix, in which at every point in the helix, the
director rotates by an angle $\theta$ toward the helix axis, and
the sample can elongate parallel to the helix axis.  A 90 degree
rotation would change the cholesteric into a nematic, with the
maximum elongation of the sample. We explore the possibility that
some smaller rotation angle will minimize the free energy as a
compromise between elongation and twisting.

\section{\label{sec:Theory}Theory}

To study the isotropic--cholesteric transition we model the gel
following the approach of Ref. \cite{Lubensky:01}. We describe the
orientational degrees of freedom  using the symmetric--traceless
nematic order parameter $Q_{ij}$, rather than a director
$\mathbf{n}$, in order to describe the development of
orientational order. The gel is modeled as an isotropic elastic
medium; its free energy can be written in terms of the right
Cauchy--Green tensor $u_{ij}$, or equivalently the left
Cauchy--Green tensor $v_{ij}$. The latter tensor transforms like a
rank--2 tensor under rotations in the target space of the elastic
medium, i.e., the space defined by the locations of the mass
points \textit{after} a distortion. The former tensor, on the
other hand, transforms like a rank--2 tensor under rotations in
the reference space, defined by the locations of the mass points
\textit{before} a distortion occurs. The nematic order parameter
tensor $Q_{ij}$ also transforms as a rank--2 tensor in the target
space. Thus, in constructing the free energy, the coupling of the
elastic degrees of freedom to the nematic degrees of freedom must
involve contractions of $Q_{ij}$ with $v_{ij}$, rather than
$u_{ij}$. The Cauchy--Green tensors are given by:
\begin{eqnarray}
u_{ij}&=& \frac{1}{2}(\partial_i u_j + \partial_j u_i + \partial_i u_k \partial_j u_k)\\
v_{ij}&=& \frac{1}{2}(\partial_i u_j + \partial_j u_i + \partial_k u_i \partial_k u_j),
\end{eqnarray}
where $\mathbf{ u(x)}$ is the displacement vector associated with
the elastic distortion, and $\partial_i$ denotes
$\partial/\partial x_i$, with $x_i$ the \textit{i}th component of
the position vector of the mass points in the reference space.
Note that $\Tr{u}=\Tr{v}$ in general, while to linear order in
$\mathbf {u(x)}$, the two elastic tensors are equal. We use the
summation convention throughout, summing over repeated indices,
which span the three--dimensional reference space.

The free energy density $f$ of the gel consists of an isotropic
elastic term $f_{el}(\utensor)$,a term $f^\prime_Q(\Q)$ describing
the nematic degrees of freedom (including gradient terms), and a
term $f_C$ which couples the elastic and nematic degrees of
freedom;
\begin{equation}
f= f_{el}(\utensor)+f^\prime_Q(\Q)+f_C
\end{equation}
The elastic energy $f_{el}(\utensor)$ is given in by:
\begin{equation}
f_{el}(\utensor)=\frac{1}{2}\lambda(\Tr{u})^2 + \mu \Tr{u}^2,
\end{equation}
where $\lambda$ and $\mu$ are the Lam\'e coefficients. It suffices
to consider only terms quadratic in $u_{ij}$.

The free energy $f^\prime_Q(\Q)$ describing the nematic degrees of
freedom is a sum of gradient \cite{Wright:89} and bulk terms:
\begin{equation}
f^\prime_Q(\Q)=\frac{1}{4}K_1(\varepsilon_{ijk}\partial_j Q_{ik} +
2 q_o Q_{ij})^2 + \frac{1}{4}K_o(\partial_j Q_{ij})^2 +
\frac{1}{2}r^\prime_Q \Tr{Q}^2 - w_3
\Tr{Q}^3+w^\prime_4(\Tr{Q}^2)^2.
\end{equation}
The elastic constants $K_o$ and $K_1$ are related to the Frank
elastic constants and the nematic order parameter $S$ by:
\begin{eqnarray}
\label{K22}K_{22}&=& K_1 S^2\\
\label{K33}K_{11}&=&K_{33}=\frac{S^2}{2}(K_o+K_1),
\end{eqnarray}
where $K_{11},K_{22}$, and $K_{33}$ are the splay, twist and bend
Frank constants respectively, $q_o$ is the cholesteric torsion of
the mesogenic molecules, and $\varepsilon_{ijk}$ is the fully
antisymmetric Levi--Civita tensor. The pitch of the ordinary
cholesteric helix would be $2\pi/q_0$.  Terms of higher order in
$Q_{ij}$ would be required to break the equality of the bend and
splay elastic constants \cite{Lubensky:70}.

The simplest coupling between the elastic and nematic degrees of
freedom is given by:
\begin{equation}
f_C=-s\Tr{u}\Tr{Q}^2 - 2t\text{Tr}{\vtr} \Q,
\end{equation}
just as in the case of the \textit{I--N} transition. Here $\vtr$
is the symmetric--traceless part of $v_{ij}$:
$\tilde{v_{ij}}=v_{ij}-\frac{1}{3}\delta_{ij}v_{kk}$. In fact, the
only difference between the energy $f$ used here and the
corresponding one used in Ref. \cite{Lubensky:01}, is the presence
in $f$ of the gradient terms proportional to $K_o$ and $K_1$,
which are required to describe cholesteric ordering.

As in Ref. \cite{Lubensky:01}, it is convenient to complete the
squares in the terms in $f$ involving $\Tr{u}$ and $\vtr$ and
write:
\begin{equation}
f=\frac{1}{2}B\lbrack\Tr{u} - (s/ B)\Tr{Q}^2\rbrack^2 +\mu
\text{Tr}\lbrack {\vtr} - (t/\mu)\Q \rbrack^2 +f_Q,
\end{equation}
where
\begin{equation}
f_Q=\frac{1}{4}K_1(\varepsilon_{ijk}\partial_j Q_{ik} + 2 q_o
Q_{ij})^2 + \frac{1}{4}K_o(\partial_j Q_{ij})^2+\frac{1}{2}r_Q
\Tr{Q}^2 - w_3 \Tr{Q}^3 +w_4(\Tr{Q}^2)^2,
\end{equation}
with $r_Q=r^\prime_Q-2(t^2/\mu)$ and $w_4=w^\prime_4-(s^2/2B)$.
Here $B$ is the bulk modulus of the gel, given in terms of the
Lam\'e coefficients by: $B=\lambda+\frac{2}{3}\mu$.

We now consider the isotropic--cholesteric transition within
mean--field theory, minimizing the free energy $f$ as a function
of $\Tr{u}$, and the independent components of the
symmetric--traceless tensors $\tilde{v}_{ij}$, and $Q_{ij}$. This
minimization requires care on two scores. First, elastic
compatibility must be ensured. The six quantities $\Tr{u},
\tilde{v}_{ij}$ must obey compatibility equations consistent with
the existence of an underlying single--valued  continuous
displacement field $\mathbf {u(x)}$. If we linearize the
Cauchy--Green elastic tensors, compatibility requires
\cite{Sokolnikoff:46}:
\begin{eqnarray}
\label{c1}
\partial_y^2(\vtw_{xx} + \onethird\Tr{u})+\partial_x^2(\vtw_{yy}
+\onethird\Tr{u})&=&2 \partial_x\partial_y \vtw_{xy}\\
\label{c2}\partial_z^2(\vtw_{yy} + \onethird\Tr{u})
+\partial_y^2(\onethird\Tr{u}-\vtw_{xx}
-\vtw_{yy})&=&2 \partial_y\partial_z \vtw_{yz}\\
\label{c3}\partial_x^2(\onethird\Tr{u}-\vtw_{xx}
-\vtw_{yy})+\partial_z^2(\vtw_{xx}
+\onethird\Tr{u})&=&2 \partial_x\partial_z \vtw_{xz}\\
\label{c4}\partial_y\partial_z(\vtw_{xx}
+\onethird\Tr{u})&=&\partial_x(-\partial_x\vtw_{yz}
+\partial_y\vtw_{xz}+\partial_z\vtw_{xy})\\
\label{c5}\partial_x\partial_z(\vtw_{yy}
+\onethird\Tr{u})&=&\partial_y(\partial_x\vtw_{yz}
-\partial_y\vtw_{xz}+\partial_z\vtw_{xy})\\
\label{c6}\partial_x\partial_y(\onethird\Tr{u}-\vtw_{xx}
-\vtw_{yy})&=&\partial_z(-\partial_x\vtw_{yz}
+\partial_y\vtw_{xz}-\partial_z\vtw_{xy}).
\end{eqnarray}

Second, minimizing with respect to the nematic degrees of freedom
requires an \textit{ansatz} for the nature of the ordering. While
gel ``blue phases" could in principle form, with local elongations
parallel to the double twist axes, here we consider the simpler
possibility of a conical helical phase, which will allow the gel
to elongate along the pitch axis (with small shears about this
axis), yet still gain twist energy from the helical ordering.
Minimization of the free energy will determine the optimal opening
angle of the cone. A conical helix also forms in ordinary
cholesterics in the presence of suitably strong (but not too
strong) magnetic fields, if $K_{22}>K_{33}$ \cite{Meyer:68}. The
qualitative similarity between a cholesteric in a magnetic field
and a cholesteric gel is not surprising, given the form of the
coupling proportional to $t$ appearing in $f_C$.

Assuming that the pitch axis in the ordered state lies along the
$z$ axis, and the director makes an angle $\theta$ with the $x-y$
plane (so that $\theta=0$ corresponds to the ordinary helix),
$Q_{ij}$ is given by:

\begin{equation}
\label{Qmatrix} \Q=S\pmatrix{\cos^2\theta\cos^2{qz}
-\onethird&\frac{1}{2}\cos^2\theta\sin {2qz} &\frac{1}{2}\sin
2\theta\cos {qz}\cr \frac{1}{2}\cos^2\theta\sin
{2qz}&\cos^2\theta\sin^2{qz}
-\onethird&\frac{1}{2}\sin2\theta\sin{qz}\cr
\frac{1}{2}\sin2\theta\cos{qz}&\frac{1}{2}\sin2\theta\sin{qz}
&\sin^2\theta-\onethird\cr},
\end{equation}

\bigskip
\noindent where $S$ is the magnitude of the nematic order parameter, and $q$
is the torsion of the conical helix in the gel. These latter two
quantities will be determined from the minimization procedure.

Within a mean--field theory treatment, there will be no dependence
on the $x$ and $y$ coordinates, in which case the compatibility
equations, Eqs. (\ref{c1}), (\ref{c4}), and (\ref{c5}) are
trivially satisfied and Eqs. (\ref{c2}), (\ref{c3}) and (\ref{c6})
reduce respectively to:
\begin{eqnarray}
\partial_z^2(\vtw_{yy}+\onethird\Tr{u})&=&0\\
\partial_z^2(\vtw_{xx}+\onethird\Tr{u})&=&0\\
\label{vxy} \partial_z^2\vtw_{xy}=0.
\end{eqnarray}

These latter compatibility equations represent integrable (i.e.
holonomic) constraints on the components of the elastic tensor. We
expect these components to be proportional to $\sin{qz}$ or
$\cos{qz}$, and thus integrating these equations produces the
constraints:
\begin{eqnarray}
\label{c11} \vtw_{xx}+\onethird\Tr{u}-C&=&0\\
\label{c22} \vtw_{yy}+\onethird\Tr{u}-C&=&0\\
\label{c33} \vtw_{xy}&=&0,
\end{eqnarray}
where $C$ is independent of $z$, and we have assumed cylindrical
symmetry about the $z$ axis. While the component $\vtw_{xy}$ could
on the basis of Eq.~(\ref{vxy}) be nonzero, minimization of $f$
along with the form of $Q_{ij}$, Eq.~(\ref{Qmatrix}),  shows that
the equilibrium value of $\vtw_{xy}$ is in fact zero.

For simplicity we consider from this point on the incompressible
limit, $B \rightarrow \infty$ (in fact, real gels are nearly
incompressible), in which case:
\begin{eqnarray}\label{incompress}
\Tr{u}&=&0\\
\label{incompress2}\vtw_{xx}&=&\vtw_{yy}=C
\end{eqnarray}

Using Eqs. (\ref{c33})--(\ref{incompress2}) we minimize $f$ with
respect to $C,\vtw_{xz},$ and $\vtw_{yz}$ and find:
\begin{eqnarray}
\label{vxx}2 C-(t/\mu)\lbrack Q_{xx}+Q_{yy}\rbrack&=&0\\
\label{vxz}2\mu\lbrack\vtw_{xz}-(t/\mu)Q_{xz}\rbrack&=&0\\
\label{vyz}2\mu\lbrack\vtw_{yz}-(t/\mu)Q_{yz}\rbrack&=&0
\end{eqnarray}

Using Eqs.~(\ref{Qmatrix}), and (\ref{vxx})--(\ref{vyz}) the free
energy $f$ in the incompressible limit reduces to:
\begin{eqnarray}
f=\frac{S^2}{2}\Biggl\lbrace\frac{1}{2}
(K_o+K_1)q^2\sin^2\theta\cos^2\theta&+&K_1(q\cos^2\theta-q_o)^2
+\frac{1}{3}K_1 q_o^2
+\frac{t^2}{\mu}\cos^4{\theta}\Biggr\rbrace \nonumber\\
& &+\frac{1}{3}r_Q S^2 - \frac{2}{9}w_3 S^3 +\frac{4}{9}w_4 S^4.
\end{eqnarray}

Minimizing $f$ with respect to $\theta$ we find:
\begin{equation}
\label{minthetaB} \frac{\partial f}{\partial \theta}=\sin{2\theta}
\Biggl\lbrace K_1  q(-q \cos^2\theta + q_o) +
\frac{1}{4}(K_o+K_1)q^2\cos{2\theta}
-(t^2/\mu)\cos^2\theta\Biggr\rbrace=0,
\end{equation}

while minimizing with respect to $q$ yields:
\begin{equation}\label{minq}
\frac{\partial f}{\partial q}=\frac{1}{2}(K_o-K_1)q\cos^2{\theta}
+K_1  q_o-\frac{1}{2}(K_o+K_1) q=0.
\end{equation}

Finally, minimizing with respect to $S$ yields:
\begin{eqnarray}
\label{SminhighB}
\frac{\partial f}{\partial S}=S\Biggl\lbrace
\frac{1}{2}(K_o+K_1)q^2\sin^2\theta\cos^2\theta
&+&K_1(q\cos^2\theta-q_o)^2
+\frac{2}{3} r_Q+(t^2/\mu)\cos^4\theta\Biggr\rbrace \nonumber\\
&-&\frac{2}{3}w_3S^2+\frac{16}{9}w_4 S^3=0.
\end{eqnarray}

Eq.~(\ref{minthetaB}) has the same form as the equation that
determines the optimal value of $\theta$ for a cholesteric with
magnetic susceptibility anisotropy $\Delta\chi$ in an effective
magnetic field $H$, with
\begin{equation}
\Delta\chi H^2 \equiv 2 (t^2S^2/\mu) \cos^2 \theta.
\end{equation}

Using Eqs.~(\ref{minq}) and (\ref{minthetaB}), the values of $q$
and $\theta$ which minimize $f$ obey:
\begin{equation}
\label{qtheta}
\frac{1}{4}(K_o+K_1) q^2=(t^2/\mu) \cos^2 \theta.
\end{equation}
In terms of the constant \begin{equation}\label{beta}
\beta=q_0\Biggl(\frac{\mu}{tS}\Biggr)\sqrt{\frac{K_{33}}{2\mu}},
\end{equation}
and using Eq.~(\ref{K33}), this relationship can be written as
\begin{equation}\label{qq0}
q=q_0\frac{\cos\theta}{\beta}.
\end{equation}
Note that $q$ and $q_0$ have the same sign.

Since the ratio of $K_{33}$ to $K_{22}$ plays an important role,
we define this ratio as $\gamma$ for the discussion below.

Using Eq.~(\ref{qq0}) to eliminate $q$, we find that the free
energy can be written as
\begin{eqnarray}\label{freenoq}
f&=&\frac{t^2S^2}{\mu\gamma}\Biggl\lbrace(1-\gamma)\cos^6\theta
+\frac{3}{2}\gamma\cos^4\theta-2\beta\cos^3\theta+\frac{4}{3}\beta^2
\Biggr\rbrace \nonumber\\
& &+\frac{1}{3}r_Q S^2 - \frac{2}{9}w_3 S^3 +\frac{4}{9}w_4 S^4,
\end{eqnarray}
and the solution to Eq.~(\ref{minthetaB}) obeys:
\begin{equation}\label{cubic}
(1-\gamma)\cos^3 \theta +\gamma \cos\theta -\beta =0,
\end{equation}
Note that $\beta$ is independent of $S$ (recall Eq.~(\ref{K33})),
so this equation is independent of $S$ and the determination of
the optimal angle $\theta$ is therefore independent of the value
of $S$, as long as nematic order exists. However, as can be seen
from Eq.~(\ref{SminhighB}), $S$ does depend on $\theta$.

\begin{figure}
\includegraphics[scale=0.5]{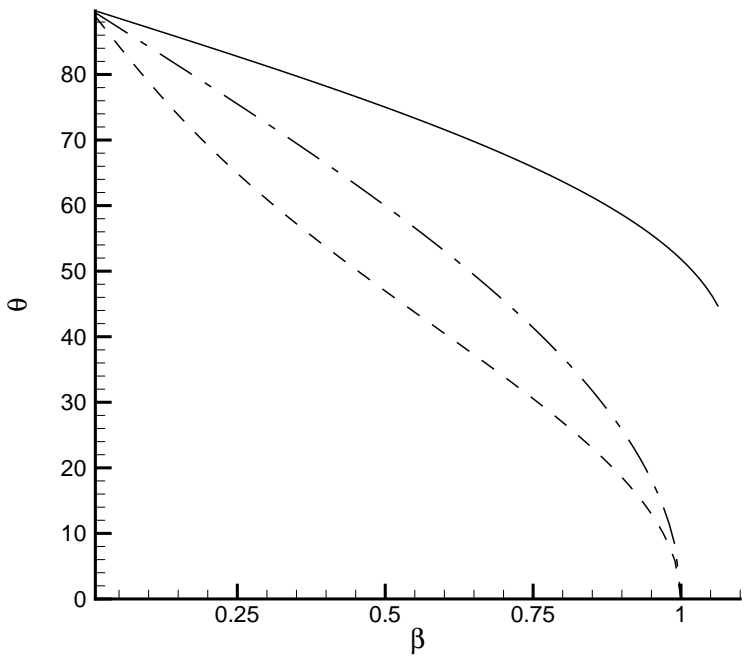}
\caption{\label{theta-beta} The optimum angle $\theta$ of the
conical helix as a function of the chirality parameter $\beta$,
for three values of the elastic constant ratio $\gamma\equiv\frac{K_{33}}{K_{22}}=2,1,$ and
0.5, for the upper, middle, and lower curves, respectively.}
\end{figure}

The constant $\beta^2$ is a measure of the chiral bending energy
in the conical helix, $K_{33}q_0^2$, measured in terms of the
coupling energy of the nematic order to the strain field in the
sample.  To picture the trade-off between spontaneous elongation
of the nematic and the chiral energy of the helix, start with
$\beta=0$; for this value, the free energy is minimized for $\theta
= \pi/2$, a simple nematic, with no helix.  For small $\beta$,
$\cos\theta \approx \beta/\gamma$, or $\theta \approx \pi/2-\beta/\gamma$, and
the helix appears with $q=q_0\cos\theta/\beta \approx q_0/\gamma$.  In
other words, even the slightest chirality produces a helix, with a
small cone angle $\phi \equiv \pi/2-\theta \approx \beta/\gamma$, and a pitch
$P \approx 2\pi\gamma/q_0$ in which bend is the dominant curvature, as
indicated by the factor $\gamma$. The factor $\gamma$ also
determines how the cone angle grows, i.e., how $\theta$ decreases,
with increasing $q_0$, or increasing $\beta$, as shown in Fig.~\ref{theta-beta}.

Real materials have positive values of $\gamma$, on the order of
2.  As seen in Fig.~\ref{theta-beta}, for values of $\gamma$ less
than about 1.5, as $\beta$ is increased to 1, $\theta$ decreases
smoothly to zero, restoring the ordinary cholesteric helix.  For
larger values of $\gamma$, as $\beta$ grows, $\theta$ decreases
smoothly, and then at some value of $\beta$ greater than 1,
$\theta$ jumps discontinuously to zero.  This is a trade-off
between bend and twist energy in the helix; since the twist
elastic constant is much less than the bend constant, the
initially bend-dominated helix can lower its energy by this
transformation.  While the helix is either gradually or suddenly
changing from bend to twist, its pitch is also evolving toward the
value determined by pure twist, as indicated by Eq.~(\ref{qq0}).

To summarize, other than the scaling of the helix energy to the
strain energy in determining the constant $\beta$, the geometric
properties of the conical helix are determined by the nematic
elastic energies.  It is interesting to estimate the strength of
the chirality needed to produce a significant cone angle.  If we
look at the case $\beta=0$, the spontaneous elongation of the
sample on entering the nematic phase is of order $tS/\mu$.  Let us
set this to a value 2.  The other factor needed in evaluating
$\beta$ for the conical helix is the length $\sqrt{K_{33}/2\mu}$.
Estimating $K_{33}=10^{-11}$ J/m, and $\mu=10^3$ J/m$^3$, for a
weak gel, this length is of order 0.1 $\mu$m. Using these numbers,
to achieve a value of $\beta$ of about 0.5, we need a helix pitch
of a few tenths of a micrometer. This pitch is common for highly
chiral nematics.

Now we turn to the nematic ordering and the elastic deformation of
the sample due to the presence of the conical helix.

Because of the form of the coupling of the nematic order to the
strain field, the energy of the conical helix only enters into a
term in the free energy proportional to $S^2$, meaning that the
chiral energy simply lowers the transition temperature of the
isotropic to cholesteric phase transition, relative to what it
would be for $q_0=0$. This effect is familiar for cholesterics in
general.

More interesting is the effect of the conical helix on the shape
of the sample.  The elastic distortion induced by the phase
transition can be determined from Eqs. (\ref{Qmatrix}),
(\ref{c33}), (\ref{incompress2}) and (\ref{vxx}), with the result
that the strain tensor $\vtensor$ is given by:

\begin{equation}
 \vtensor=\pmatrix{\frac{tS}{2\mu}(\cos^2\theta - \frac{2}{3})&0&\frac{tS}{2\mu}\sin{2\theta}\cos{qz}\cr
0&\frac{tS}{2\mu}(\cos^2\theta -
\frac{2}{3})&\frac{tS}{2\mu}\sin{2\theta}\sin{qz}\cr
\frac{tS}{2\mu}\sin{2\theta}\cos{qz}&\frac{tS}{2\mu}\sin{2\theta}\sin{qz}&\frac{tS}{\mu}(\frac{2}{3}-\cos^2\theta)\cr}
\end{equation}

\bigskip
This equation can be easily solved for the displacement field
$\mathbf{u}$, consistent with the compatibility requirements and
incompressibility, with the results:
\begin{eqnarray}
\label{ux}u_x&=&\frac{tS}{2 \mu}(\cos^2\theta -2/3)x
+ \frac{tS}{\mu}\sin{2\theta}\sin{qz} \\
\label{uy}u_y&=&\frac{tS}{2 \mu}(\cos^2\theta -2/3)y
- \frac{tS}{\mu}\sin{2\theta}\cos{qz} \\
\label{uz}u_z&=&-\frac{tS}{\mu}(\cos^2\theta -2/3)z
\end{eqnarray}

\begin{figure}
\includegraphics[scale=0.5]{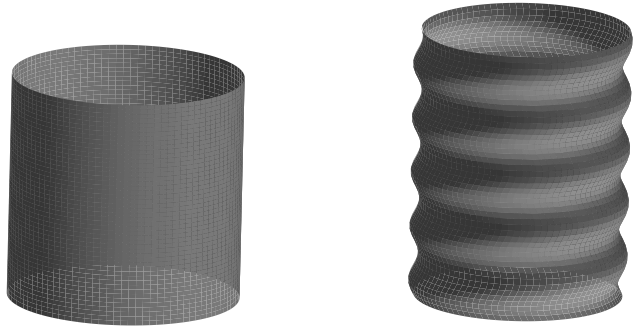}
\caption{\label{small} The distortion of a cylindrical isotropic
gel (left) after undergoing a transition to the conical helix
phase (right) with a small cone angle.}
\end{figure}

\begin{figure}
\includegraphics[scale=0.5]{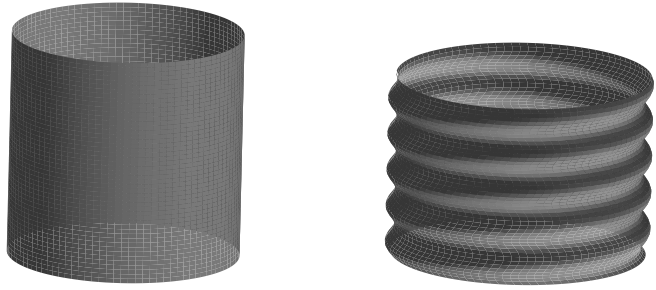}
\caption{\label{large} The distortion of a cylindrical
isotropic gel (left) after undergoing a transition to the conical
helix phase (right) with a large cone angle.}
\end{figure}

We illustrate this displacement field in Figs.~\ref{small} and
\ref{large} where we show the deformation of a cylindrical gel in
the isotropic phase which then undergoes a transition to the
cholesteric phase.
As anticipated, for a small cone angle $\phi=\pi/2-\theta$, such
that $\cos^2\theta<2/3$, the sample is elongated parallel to the
helix axis, and the transverse shears produce ridges on its
surface, as shown in Fig.~\ref{small}.  For large cone angle,
i.e., for $\cos^2\theta>2/3$, the helix is close to its twist
form, and in fact the sample has shrunk along the helix axis, and
expanded laterally, again with ridges, as shown in
Fig.~\ref{large}. This unexpected deformation represents a
compromise in which again there is elongation parallel to the
local director in the helix, but elastic compatibility demands
equal elongation perpendicular to both the director and the helix
axis, with shrinkage parallel to the helix axis to maintain
constant volume.  The possibility of this mode of deformation was
first pointed out to us by Mark Warner \cite{Warner}.

\section{\label{sec:Conclusions}Conclusions}

In conclusion, we have found that the conical helix is a possible
free energy minimizing texture of the cholesteric gel phase.  It
allows for a combination of local elongation of the system
parallel to the nematic director, with some twisting due to the
chirality of the material.  It is fascinating to speculate on the
existence of more complex textures that may offer an even better
accommodation of the competing tendencies for spontaneous
elongation and twist. We are currently looking into the
possibility of periodic textures, similar to the blue phases,
which may play this role.

\section*{Acknowledgments}

We thank T.~C. Lubensky and M.~Warner for helpful discussions, and
G. Loriot for assistance with Figs.~\ref{small} and \ref{large}. This work was
supported by the National Science Foundation under Grant Nos.
DMR--9873849, DMR--9974388, and DMR--0131573.

\end{document}